\documentclass[epjCONF, onecolumn]{svjour}
\usepackage{graphics}
\usepackage[varg]{txfonts} 
\usepackage[latin1]{inputenc}
\session-title{Assembling the Puzzle of the Milky Way}
\begin{document}
\title{Counter-Orbiting Tidal Debris as the Origin of the MW DoS}
\author{Marcel S. Pawlowski\inst{1}\fnmsep\thanks{\email{mpawlow@astro.uni-bonn.de}}}
\institute{Argelander Institute of Astronomy, Auf dem H\"ugel 71, D-53121 Bonn, Germany}
\abstract{
The Milky Way satellite galaxies show a phase-space distribution that is not expected from the standard scenario of galaxy formation. This is a strong hint at them being of tidal origin, which would naturally explain their spacial distribution in a disc of satellites. It is shown that also their orbital directions can be reproduced with the debris of galaxy collisions. Both co- and counter-orbiting satellites are formed naturally in merger and fly-by interactions.
} 
\maketitle
\section{Introduction}
\label{intro}

The ''classical'' Milky Way (MW) satellite galaxies define a disc nearly perpendicular to the MW plane \cite{Metz2007}. Similarly, the ''faint'' satellites, mostly detected in the SDSS, independently define the same plane \cite{Kroupa2010}. Analysing the young halo globular clusters in a similar manner, it is found that they define a similar disc, too \cite{Pawlowski2011b}. This disc of satellites (DoS) can be described by the normal vector perpendicular it.

From the known proper motions of MW satellite galaxies their orbital poles (direction of angular momentum vector) can be determined \cite{Metz2008}. It is found that six out of eight satellites are in agreement with co-orbiting within the DoS. The Sculptor satellite galaxy is counter-orbiting, its orbital pole is $\sim 180^\circ$ off, making that satellite orbit within the DoS, but in the opposite direction. The average orbital pole of the six best-aligned satellites lies close to the DoS normal, suggesting that the DoS is rotationally supported.

The spacial and orbital alignment is extremely unlikely if the MW satellites are of cosmological origin. The alternative scenario suggest that they are tidal dwarf galaxies (TDGs). 
In this scenario, material was tidally expelled during an early galaxy-interaction (about 10 Gyr ago). This tidal debris is naturally aligned in the plane of the interaction. TDGs form from this debris, occupying the same phase-space region. 
Using stellar-dynamical models of galaxy interactions it is tested whether both co- and counter-orbiting tidal debris form in single merger and fly-by interactions. Thus, whether debris pro- and retrograde with respect to the interaction orbit can be found.
The calculations use the particle-mesh code {\scshape{Superbox}\footnotesize\raise.4ex\hbox{+\kern-.2em+}}. A parameter scan including interactions with 1-to-1 and 4-to-1 mass ratios of target-to-infalling galaxy is performed, resulting in a total of 74 models, taking more than 200 CPU-days.

\section{Result}
\label{result}
\begin{figure}
\center
\resizebox{0.80\columnwidth}{!}{
  \includegraphics{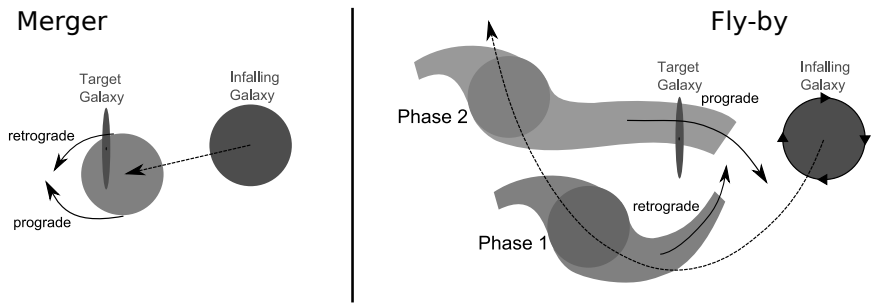} }
\caption{Sketch of the interaction geometry of the models and the resulting origin of pro- and retrograde material in galaxy interactions. The infalling galaxy, approaching from the right, is seen face-on. The target galaxy is seen edge-on. Tidal debris forms a disc in the plane of the sketch.}
\label{fig:sketch}       
\end{figure}

The interaction geometry and the typical origins of pro- and retrograde tidal debris are sketched in Figure \ref{fig:sketch}. Co- and counter-orbiting tidal material forms naturally in single galaxy interactions, in both mergers and fly-bys. It resembles the observed orbital pole distribution of MW satellite galaxies \cite{Pawlowski2011}.

In the case of galaxy-mergers, pro- and retrograde material forms readily, the radial distributions for both orbital directions are similar. The fraction of counter-orbiting material varies between a few and up to 50 per cent.

In fly-by interactions, the same two-phase origin is found in all models. It is schematically sketched in the right panel of Figure \ref{fig:sketch}. After the galaxies pass, a tidal tail forms, along which particles stream towards the target galaxy. This way, retrograde orbits form first. Then the tidal tail sweeps over the center of the target galaxy, changing the orbital direction of subsequently accreted particles and thus leading to prograde orbits. This two-phase origin gives a maximum radius up to which counter-orbiting material can be expected. In most model calculations, the majority of tidal debris with high apogalactica is prograde, retrograde particles usually account for a few up to 20 per cent, but sometimes up to almost 100 per cent.


The tidal material resulting from the modelled interactions resembles the DoS around the MW. The debris orbit within a disc defined by the chosen interaction geometry, leading to a polar disc with co- and counter-orbiting material. TDGs will form from the tidal debris and therefore occupy the phase-space region of the debris. This scenario explains the peculiarities in both the spatial and orbital distribution of the MW satellite system. It might even allow to constrain the early MW interaction that shaped the satellite distribution, as it is found that faster relative velocities of the colliding galaxies lead to higher fractions of retrograde material.

One possible progenitor for an infalling galaxy in a fly-by interaction was already suggested by Lynden-Bell \cite{Lynden-Bell1976}: a Magellan Clouds progenitor galaxy might be the parent of the MW satellite galaxies. Compared to our results the position and orbit of the LMC, which lies and orbits within the DoS, agrees with this scenario. Furthermore, the Magellanic Stream is close to the DoS \cite{Pawlowski2011b}. The fly-by models predict that the majority (defining the co-orbiting direction) of tidal debris and thus most probably satellite galaxies would be prograde with respect to the infalling galaxy. This is the case for the MW satellites, as the majority have orbital poles close to the LMC pole. Furthermore, the counter-orbiting satellite galaxy, Sculptor, has a relatively low apocenter, which agrees with the maximum radius found for retrograde material in the two-phase origin of fly-by interactions.

The presented results show that the scenario of a tidal origin of the MW DoS is very promising and in fact the most natural explanation for the DoS. More detailed model calculations, including gas physics and the formation of TDGs will be needed in order to reconstruct in detail the early encounter that shaped the MW and created its satellite galaxies.

%
%

\end{document}